\documentclass{article}
\usepackage[utf8]{inputenc}
\usepackage{caption}
\usepackage{subcaption}
\usepackage{multicol}
\usepackage{xcolor}

\usepackage{cite}

\newcommand{\sqrts}{\mbox{$\sqrt{\mathrm{s}}$}}

\newcommand{\lam}{$\Lambda$}

\newcommand{\ppt}{$p_{\rm T}$}

\usepackage{graphicx}
\providecommand{\keywords}[1]{\textbf{\textit{Keywords---}} #1}
\usepackage{amssymb}

\usepackage{graphicx}
\usepackage{multicol}

\usepackage{lineno}
\begin{document}


\begin{center}
{\Large \bf A Comprehensive Study of Energy Dependence of Particle Ratios in $pp$ Collisions from SPS to LHC Energies }

\vskip1.0cm

A.~M.~Khan$^{1}$,
M.~U.~Ashraf$^{2}${\footnote{Email: usman.ashraf@cern.ch}}
Junaid~Tariq$^{3}$,
Anwarzada$^{4}$, 
Ijaz Ahmed$^{4}$

{\small\it 
$^1$Key Laboratory of Quark \& Lepton Physics (MOE) and Institute of Particle Physics,
Central China Normal University, Wuhan 430079, China\\
$^2$ Pakistan Institute of Nuclear Science and Technology (PINSTECH), Islamabad 45650, Pakistan\\
$^3$ Department of Physics, Quaid-i-Azam University, Islamabad 44000, Pakistan\\
$^4$ Department of Physics, Riphah International University, Islamabad 44000, Pakistan\\

}
\end{center}

\vskip1.0cm


\begin{abstract}

A comprehensive study has been performed to estimate the kaon to pion ($K^\pm$/$\pi^\pm$) yield ratio and total kaon to total pion ($K$/$\pi$) yield ratio as a function of centre of mass energy in $pp$ collisions at different energies i.e., \sqrts~= 6.3, 17.3, 62.4, 200, 900 GeV, 2.76 TeV, 7 TeV, 13 TeV and 14 TeV using EPOS1.99, EPOS-LHC, HIJING, QGSJETII-03 and Sibyll2.3d model simulations. NA61/SHINE experiment reported that $K^+/\pi^+$ yield ratio exhibits rapid changes at the SPS energy range. A horn like structure appears in $K/\pi$ yield ratio as a function of collision energy. Significant presence of horn in $K^+$/$\pi^+$ and $K^-$/$\pi^-$ yield ratio is suggested by experimental data at lower energies, which is confirmed by HIJING and EPOS-LHC models. A smooth increase in $K/\pi$ yield ratio is also seen at higher energies. The model simulations predict similar increase in yield ratio with increasing energies. On the basis of previous available measurements, we also study model predictions of these yield ratios at \sqrts~= 13 and 14 TeV where no data is available. Almost all models suggest a saturation in the yield ratio within statistical fluctuations at these energies except EPOS1.99 and QGSJETII-03 which slightly over predict the yield ratio. These systematic comparisons are helpful to apply possible constraints on various hadronic event generators to significantly improve the predictions of Standard Model physics as well as for the understanding of underlying physics mechanisms in high energy collisions.

\vskip0.5cm
\keywords{Hadronic event generators, $pp$ collisions, Monte-Carlo Simulation, Prediction, LHC energies}
\end{abstract}


\section{Introduction}\label{sec1}


The nucleus-nucleus ($AA$) collisions at ultra-relativistic energies has been one of the major area of interest for experimental and theoretical physicists. $AA$ collisions could be helpful to extract information about the spatiotemporal evolution of multi-particle production processes, one of the primary interests in view of recent progresses in Quantum Chromodynamics (QCD). In addition to $AA$ interactions, the study of proton-proton ($pp$) collisions are also important because it provides input to theoretical models based on strong interactions. Another important aspect is that, it acts as a baseline to understand the $AA$ collisions at relativistic and ultra-relativistic energies. This reference is also needed for the investigation of possible initial state effects in the collisions. The production of soft particle in $pp$ interactions is sensitive to the hadronization of quark, flavor distribution inside the proton and baryon number transport. 

The multiplicity distribution of produced particles in $pp$ collisions is one of the basic observables which shows the properties of underlying production mechanisms of different particles. In addition to production of $\pi^\pm$, the production of $K^\pm$ is also of great interest due to the fact that strangeness production is a sensitive probe to study the hadronic interactions as well as hadronization in $pp$ and ultra-relativistic $AA$ collisions. The strangeness enhancement in $AA$ collisions is suggested as a possible signature of quark-gluon plasma (QGP) ~\cite{5}. It is a fact that initial state collisions does not contain any strange quark or strange anti-quark but production of kaons confirms that the strange quark and anti-strange quark pair is created during the collisions between nucleons and nuclei. The nuclei collisions form a high energy density fireball which expand rapidly and a partonic phase of quasi free quarks and gluons, the QGP is expected to produce~\cite{6}. Therefore, investigation of these interactions may provide useful information to distinguish between hadronic and partonic matter as well as properties of phase transition in between.

The production of $K^\pm$ will shed light to understand the strangeness production mechanisms in elementary ($pp$) collisions~\cite{5}. In high energy collisions, $K/\pi$ ratio is suggested as a key tool to study the strongly interacting matter by means of relative strangeness yield. Another important aspect to study this ratio is to address the possible questions regarding the phase transition but also helps to better understand the hadronization and pre-equilibrium dynamics of the system. The data from NA61/SHINE experiment reflect rapid changes in $K^+/\pi^+$ yield ratio at the SPS energy range~\cite{NA49:2002pzu, NA49:2007stj, Pulawski:2015tka}. A horn like structure appears in $K/\pi$ yield ratio as a function of collision energy as reported in~\cite{NA49:2002pzu, NA49:2007stj, Pulawski:2015tka}. Similar structure is also observed in $AA$ collisions around \sqrts~$\approx$~8 GeV, where one expects the transition between confined and deconfined matter with the creation of mixed phases, which may indicate the onset of deconfinement in comparison with smaller colliding systems~\cite{Gazdzicki:2010iv, 7}. The significant difference is reported in the ratio upto \sqrts~= 200 GeV while at higher energies this difference becomes insignificant. This difference may arise due to the underlying production mechanisms of $K^+$ and $K^-$ at lower and higher energies. Two possible mechanisms are involved to study the $K^\pm$ production, first is pair production which is dominated at higher energies and the other is associated production which is dominated at lower energies.

This paper presents a comparison of $K^+/\pi^+$, $K^-/\pi^-$ and $K/\pi$ yield ratios from experimental data at different energies, i.e., \sqrts~= 6.3, 17.3, 62.4, 200, 900 GeV, 2.76 TeV, 13 TeV and 14 TeV in $pp$ collisions with EPOS1.99, EPOS-LHC, HIJING, QGSJETII-03 and Sibyll2.3d model simulations. The paper is organized as follows; Brief introduction of the models is presented in section~\ref{sec2}. In section~\ref{sec3} results and discussion is presented followed by conclusion in section~\ref{sec4}.

\section{Model Details}\label{sec2}
 It is not yet possible to perform calculations with first principles of Quantum Chromodynamics (QCD) for the observables related to bulk of the produced particles at colliders in high energy interactions. Phenomenological models, relying on basic principles of Quantum Field Theory (QFT) and predictions of pQCD together with phenomenological fits, are instead used for the predictions of various observables high energy interactions~\cite{dEnterria:2011twh}. The various models used for comparisons are briefly discussed in the remaining part of this section.

{\bf DPMJET}~\cite{Roesler:2000he} is based on Dual Parton Model (DPM) for the description of soft and multi-particle interactions in high-energy collisions. Soft processes are described by pomerons exchange under the Regge theory scheme and hard processes by using perturbative parton scattering approach. DPMJET works on the principles of multiple scattering Gribov-Glauber formalism and can be used to simulate a wide range of $hh$, $\gamma h$, $\gamma \gamma$, $AB$ and $\gamma A$ collisions for energies ranging from few GeV to cosmic-ray interactions of the highest energy scale. The physics models and flexibility of DPMJET allows for the calculations of total, (quasi) elastic as well as production cross-sections for various colliding systems at high energies~\cite{Roesler:2000he}. The hadronic interaction model for $pp$ collisions in the DPMJET is derived from PHOJET while the fragmentation configurations are acceded from PYTHIA Lund model~\cite{ATLAS:2020bhl}. DPMJETIII used for this study integrated the features of PHOJET, DPMJETII and DTUNUC2 models with Glauber-Gribov calculations for intra-nuclear cascades, excited nuclei and various nuclear cross-sections~\cite{Roesler:2000he}. An account of DPMJET model characteristics and upgraded features is made in Refs.~\cite{Roesler:2000he, Bopp:2005cr}. 

{\bf EPOS}~\cite{Pierog:2009zt} is based on a scattering approach in which partons and strings are treated consistently in a quantum mechanical framework. A high-energy hadronic interaction, in simple parton based models is considered as a “parton ladder” exchange between participants of the interaction. The “parton ladder” in EPOS has two parts, a hard-scattering part and an entirely phenomenological soft part parameterized in Regge pole fashion. Therefore, it is based on perturbative QCD, Gribov-Regge multiple scattering, and string fragmentation. EPOS-LHC model has same theoretical foundation as the EPOS1.99. EPOS-LHC makes a few adjustments to the parameters related to flow of the high-density core of thermalized matter created following $pp$ or $AA$ collisions~\cite{Pierog:2013ria}. This model is implemented for the hadronization phase as modifications to the string fragmentation depending on string density in final state. Particularly, it is important to note that this model gets the relation between mean transverse momentum and charged particle multiplicity without implementation of color reconnection in comparison with models like Pythia~\cite{Sjostrand:2007gs} and Herwig~\cite{Bellm:2015jjp}. In addition, EPOS includes off-shell remnants in the picture and thus can solve multi-strange baryon problem arising in the conventional interaction models. Effects related to consistent cross-section calculation with energy conservation, Cronin transverse momentum broadening, parton saturation, screening and collective behaviour in heavy-ion interactions have also been included in EPOS. The version EPOS1.99 included the non-linear effects, reduced cross section and inelasticity as compared to its older versions. Updated versions of EPOS-LHC and EPOS1.99 are used forcurrent study. More details on EPOS model and developments included in EPOS1.99 can be found in Ref.~\cite{Pierog:2009zt}.

{ \bf QGSJET}~\cite{Ostapchenko:2007qb} hadronic interaction model is developed in the framework of Quark-Gluon String model~\cite{Engel:2011zzb}. The description of semi-hard processes as “semi-hard pomeron” approach and a scheme for incorporating the heavy-ion interactions were later included in the model~\cite{Kalmykov:1993qe}. In QGSJET, a scattering is considered as a pomeron exchange process having two different (soft and semi-hard) components. Nucleus-Nucleus or hadron-hadron interactions are then modelled under the Gribove’s Reggeon theory as multiple scattering, but there is no fluid component and Lund algorithm is used to disintegrate supercritical pomerons into strings. Abramovski Gribov Kancheli (AGK) cutting rules and optical theorem are employed to estimate the cross sections of final states. Parton cascade overlapping at higher energies or in central collisions create significant nonlinear effects in the interactions, these effects are described as pomeron-pomeron interactions in the Raggeon Field Theory (RFT). The version QGSJETII-03 used for this study has been designed to include the nonlinear effects at the fundamental level by enhanced pomeron diagrams approach~\cite{Ostapchenko:2004ss}. Further details about the QGSJET model and QGSJETII can be seen in Refs.~\cite{Ostapchenko:2007qb, Ostapchenko:2004ss, Engel:2011zzb, Kalmykov:1993qe}.

{\bf HIJING}~\cite{Wang:1991hta} model -- an acronym for heavy-ion jet interaction generator. The pQCD inspired model, Dual Parton model (DPM)~\cite{Capella:1979fm} for the study of jet fragmentation, while to study the effects of soft interactions at low and medium energies it uses Lund fragmentation~\cite{Andersson:2001yu}. The particular development of this model is related to study the parton distribution functions (PDF), associated production of particles, jets and mini-jets produced in dense medium and soft excitation processes~\cite{Wang:1991hta}. HIJING can simulate multi-particles production in different systems i.e., $pp$ and $AA$ up to energy range of $\sqrt{s}= 5-2000 $~GeV~\cite{Capella:1979fm, Wang:1991hta}. At the time of development, HIJING was the only model incorporates the pQCD methodology of multiple-jet processes from Pythia and other related processes including parton shadowing and jet quenching.

{ \bf Sibyll}~\cite{Riehn:2019jet} event generator describes the small angle production and projectile direction flow very well as it was designed majorly to understand the air showers and cosmic ray interactions in the Earth’s atmosphere. The interaction aspects related to jet production at higher {\ppt} and electroweak processes are not very well embedded in the workings of the model~\cite{Riehn:2019jet}. However, implementation of basic principles from unitarity and scattering theory empowers Sibyll to be used for phase space and energies of the interactions that are beyond the scope of modern colliders~\cite{Riehn:2017mfm}. Nonetheless, Sibyll has been used to well reproduce the LHC Run-I data~\cite{Riehn:2019jet}. The upgraded version Sibyll2.3 included better fits describing the $p/\pi/K$ elastic and total cross sections with inputs from experimental data. Improvement in the modeling of fragmentation region is incorporated in Sibyll2.3c. The improved version, Sibyll2.3d used for this study gives better $\pi^{+}/\pi^{0}$ ratios description along with other features that are important for the hadronization mechanism and production of muons in extensive air-showers. Refs.~\cite{Riehn:2019jet, Riehn:2017mfm, CMS:2015zrm} provide further details of the Sibyll model and its upgraded versions. 

\section{Analysis, Results and Discussion}\label{sec3}

For the current study, 0.1 million events have been generated using HIJING, EPOS1.99, EPOS-LHC, QGSJETII-03 and Sibyll2.3d at various beam energies i.e., \sqrts~= 6.3, 17.3, 62.4, 200, 900 GeV and 2.76 TeV, 7 TeV, 13 TeV and 14 TeV in $pp$ collisions to study the energy dependence of $K^+/\pi^+$, $K^-/\pi^-$ and $K/\pi$ yield ratios. We have compared our results with the published data available so far i.e., from \sqrts~= 6.3 GeV upto 7 TeV. None of the experiment at LHC reported these ratios at energy \sqrts~ $>$ 7 TeV. LHC has started high luminosity Run-III data taking at \sqrts~= 13.6 TeV this year after more than three years of maintenance. Therefore, it is important to study the model predictions of these ratios at higher LHC energies to apply possible constraints on various hadronic event generators to improve the standard model physics predictions as well as underlying physics mechanisms in high energy collisions. On the basis of previous published results at various energies, we also study the predictions of various models of these yield ratios at \sqrts~= 13 and 14 TeV where no data is available so far.

\subsection{$K^+/\pi^+$ Ratio}

The results of $K^+/\pi^+$ ratio measured by various experiments at \sqrts~= 6.3 GeV~\cite{Pulawski:2015tka}, 17.3 GeV~\cite{NA49:2009brx}, 62.4 GeV~\cite{PHENIX:2011rvu}, 200 GeV~\cite{STAR:2008med}, 900 GeV~\cite{ALICE:2011gmo} and \sqrts~ = 2.76 TeV~\cite{ALICE:2015ial}, 7 TeV~\cite{ALICE:2015ial} is listed in table 1. The experimental results of $K^+/\pi^+$ ratios from inelastic $pp$ collisions \sqrts~= 6.3 GeV at mid rapidity from NA61/SHINE Collaboration is presented in Ref.~\cite{Pulawski:2015tka}. It has been reported that, a rapid changes in the energy dependence of $K^+/\pi^+$ ratios is observed in the SPS energy regime~\cite{Pulawski:2015tka}. 

\begin{figure}[ht!]
\begin{center}
\includegraphics[width=0.8\textwidth]{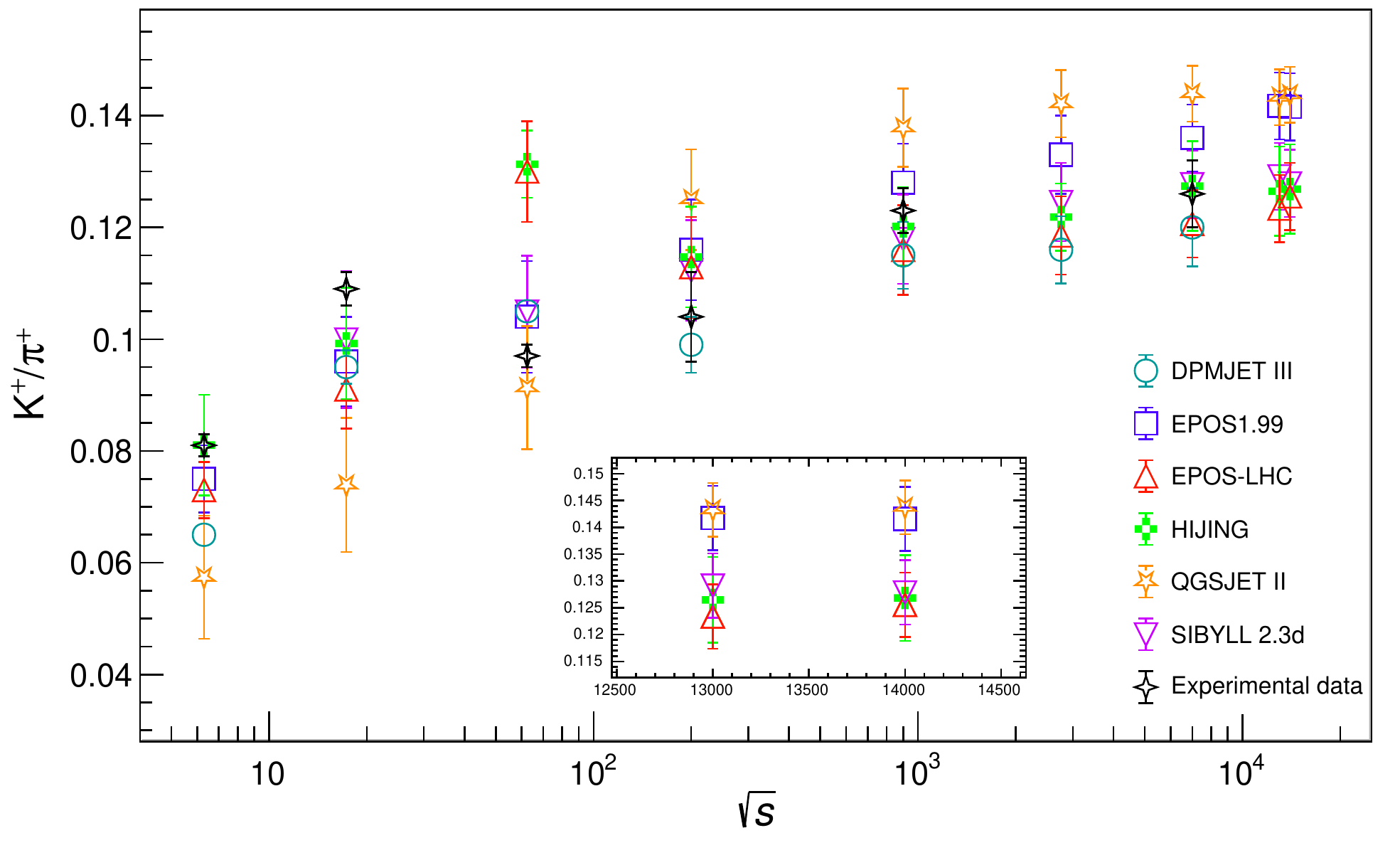}
\caption{ Energy dependence of $K^+/\pi^+$ yield ratio at \sqrts~= 6.3 GeV upto \sqrts~= 14 TeV in $pp$ collisions from DPMJETIII, EPOS1.99, EPOS-LHC, HIJING, QGSJETII-03 and Sibyll2.3d compared with the experimental data (where available). The model predictions at \sqrts~ = 13 and 14 TeV are shown in sub-pad for better visualization.}
\label{fig1}
\end{center}
\end{figure}

Figure~\ref{fig1} shows $K^+/\pi^+$ yield ratio as a function of centre of mass energy (\sqrts) in $pp$ collisions from EPOS1.99, EPOS-LHC, HIJING and Sibyll2.3d results in comparison with published experimental measurements. It has been observed that $K^+/\pi^+$ ratios in case of DPMJETIII simulations show increasing trend upto \sqrts~= 62.4 GeV and sudden decrease at \sqrts~= 200 GeV then start to increase again with increasing energy. There is no prominent saturation is seen in case of DPMJETIII towards LHC energy regime. The DPMJETIII simulations confirm the presence of horn, which is seen in the experimental measurements. The DPMJETIII simulations data points for $K^+/\pi^+$ ratio is taken from Ref.~\cite{Bhattacharyya:2017rmc}. $K^+/\pi^+$ ratio extracted from the EPOS1.99 model increases with the increase in energy and start to saturate at \sqrts~ $\ge$ 13 TeV. While different scenario has been observed in case of EPOS-LHC. The $K^+/\pi^+$ ratio from EPOS-LHC increases upto \sqrts~= 62.4 GeV and shows a sudden decrease at \sqrts~= 200 GeV which start to increase again at higher energies indicating the horn. EPOS-LHC clearly over predict the experimental data at \sqrts~= 62.4 GeV. In case of HIJING, the ratio increases upto \sqrts~= 62.4 GeV, decrease at \sqrts~= 200 GeV and increases again at higher energies reflects the presence of horn. QGSJETII-03 and Sibyll2.3d on the other hand, shows smooth increasing trend of ratios with increasing energy. It has also been observed that Sibyll2.3d does not produce the simulations at \sqrts~ $<$ 10 GeV, therefore there is no comparison of experimental data with Sibyll2.3d at \sqrts~= 6.3 GeV. However, the abrupt increase in the ratio is seen in case of QGSJETII-03 in between \sqrts~= 62.4 GeV and \sqrts~= 200 GeV and the ratio saturate at higher energies. EPOS1.99, HIJING and Sibyll2.3d results reasonably reproduce the experimental measurements within statistical fluctuations. The sudden decrease in $K^+/\pi^+$ ratio is not confirmed by EPOS1.99 and Sibyll2.3d. However, large error bars in EPOS1.99 at \sqrts~ $\le$ 62.4 GeV make it difficult to claim the presence of horn in the ratio. The experimental measurements also shows significant presence of horn in the yield ratio. The predictions of these yield ratios with various models at \sqrts~ = 13 and 14 TeV where no experimental data is available so far are also presented and shown in the sub-figure~\ref{fig1} for better visualization. There is no prediction reported in Ref.~\cite{Bhattacharyya:2017rmc} from DPMJETIII model of these ratios at \sqrts~ =~13 and 14 TeV. The predictions of EPOS1.99 and QGSJETII-03 are higher as compared to the other models. However, EPOS-LHC, Sibyll2.3d and HIJING model predictions show saturation in the yield ratios at \sqrts~ = 13 and 14 TeV.

\begin{table*}[!htb]\label{tab1}
{
\begin{center}
\caption{$K^+/\pi^+$ ratio at different energies, i.e., $\sqrt{s}$ = 6.3 GeV to 14 TeV by using simulations from various MC models along with the comparison of experimental values are displayed here.}
\begin{tabular}{ccccccccccc}\\ \hline\hline

 \small Energy& \small DPMJET-III&\small EPOS1.99 &\small EPOS-LHC  &\small HIJING &\small Sibyll2.3d&\small QGSJET II &\small Experimental \\
 $\sqrt{s}$  &      &   &    &   & & \\
\hline
    \small 6.3  &\small 0.065$\pm$ 0.002 &\small 0.075$\pm$0.006 &\small 0.073$\pm$0.005 &\small 0.081 $\pm$0.009 &\small ---- & \small 0.057$\pm$0.008&\small 0.081$\pm$0.002\\
    \small 17.3 & \small  0.095$\pm$ 0.003  & \small 0.096$\pm$0.008 &\small 0.091$\pm$0.007 & \small 0.099$\pm$0.010 & \small0.100$\pm$0.0122 &\small 0.074$\pm$0.009 &\small 0.109$\pm$0.003\\
    \small 62.4 &\small 0.105$\pm$  0.002 &\small 0.104$\pm$0.010 &\small 0.130$\pm$0.009 &\small 0.131$\pm$0.006 &\small 0.105$\pm$0.010 &\small 0.091$\pm$0.008 &\small 0.097$\pm$0.002\\
    \small 200 &\small 0.099$\pm$0.005 &\small 0.116$\pm$0.009 &\small 0.113$\pm$0.009 &\small  0.115$\pm$0.009&\small 0.112$\pm$0.009 &\small 0.125$\pm$0.009 &\small 0.104$\pm$0.008\\
    \small 900 &\small  0.115$\pm$ 0.006 &\small 0.128$\pm$0.007 &\small 0.116$\pm$0.008 &\small 0.120$\pm$0.007 &\small 0.118 $\pm$0.008 &\small 0.137$\pm$0.007 &\small 0.123$\pm$0.004 \\
    \small 2760 &\small 0.116$\pm$  0.006 &\small 0.133$\pm$0.007 &\small 0.119$\pm$0.007 &\small 0.122$\pm$0.006 &\small 0.125$\pm$0.007 &\small 0.142$\pm$0.006 &\small ----\\
    \small 7000 &\small  0.120$\pm$ 0.007 &\small 0.137$\pm$0.006 &\small 0.121$\pm$0.006 &\small 0.125$\pm$0.008 &\small 0.128$\pm$0.006 &\small 0.144$\pm$0.005&\small  0.126$\pm$0.006\\
    \small 13000  &\small ---- &\small 0.142$\pm$0.006 &\small 0.123$\pm$0.006 &\small 0.126$\pm$0.008 &\small 0.129  $\pm$0.006 &\small 0.143$\pm$0.005& \small----\\
    \small 14000  &\small---- &\small 0.142$\pm$0.006 &\small 0.126$\pm$0.006 &\small 0.127$\pm$0.008 &\small  0.128$\pm$0.006 &\small 0.143   $\pm$0.005 &\small ----\\
\hline
\end{tabular}%
\end{center}}
\end{table*}


\subsection{$K^-/\pi^-$ Ratio}\label{subsec1}

\begin{figure}[ht!]
\begin{center}
\includegraphics[width=0.8\textwidth]{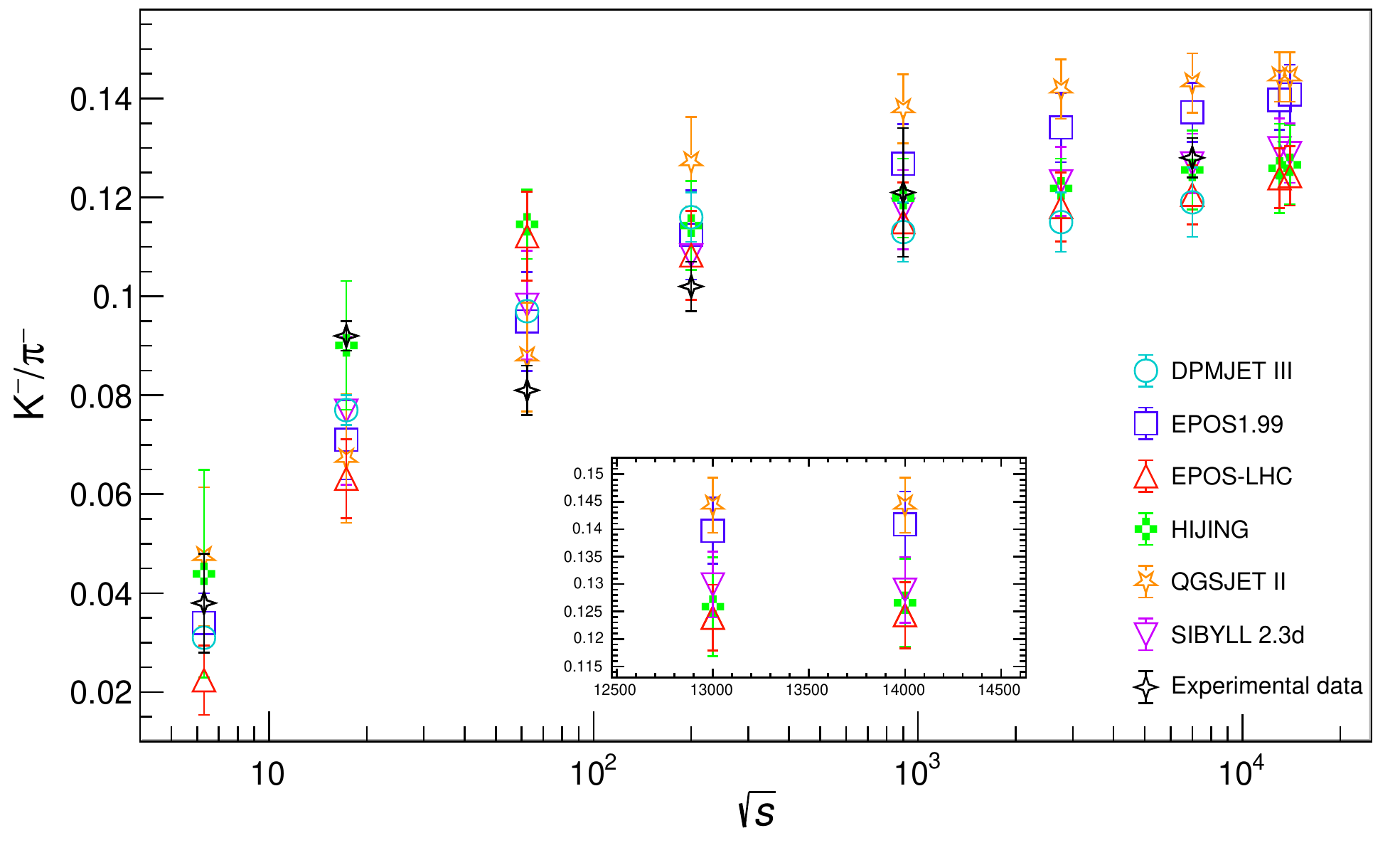}
\caption{ Energy dependence of $K^-/\pi^-$ yield ratio at \sqrts~= 6.3 GeV upto \sqrts~= 14 TeV in $pp$ collisions from DPMJETIII, EPOS1.99, EPOS-LHC, HIJING, QGSJETII-03 and Sibyll2.3d compared with the experimental data (where available). The model predictions at \sqrts~ = 13 and 14 TeV are shown in sub-pad for better visualization.}

\label{fig2}
\end{center}
\end{figure}

The energy dependence of $K^-/\pi^-$ ratio is computed using HIJING, SIbyll2.3d, EPOS-LHC, DPMEJETIII and QGSJETII-03 model simulations at \sqrts~= 6.3 GeV -- 14 TeV in $pp$ collisions. 
Figure~\ref{fig2} presents $K^-/\pi^-$ ratio as a function of energy from various model simulations in comparison with the experimental data. At lower energies \sqrts~ $\le$ 200 GeV, the DPMJETIII value is increasing with energy and drop suddenly at \sqrts~= 900 GeV and starts to increase again at higher energies showing the presence of horn as observed in the energy dependence of $K^+/\pi^+$ yield ratios~\cite{Bhattacharyya:2017rmc}. In case of HIJING and EPOS-LHC the ratio increases with energy upto \sqrts~ $\le$ 62.4 GeV and decreases suddenly at \sqrts~= 200 GeV and start to increase again at higher energies which confirms the experimental claim of the presence of horn~\cite{NA49:2002pzu, NA49:2007stj, Pulawski:2015tka}. It has also been observed that at \sqrts~= 62.4 GeV, HIJING and EPOS-LHC clearly over predict the experimental results while a nice comparison at rest of energies. The ratio from EPOS1.99 and Sibyll2.3d models increases smoothly with the increase in energy and fails to shows the horn structure. On the other hand, QGSJETII-03 values of ratio smoothly increase upto \sqrts~ $\le$ 62.4 GeV and shows a sudden increase which continues towards higher energies. The predictions of $K^-/\pi^-$ yield ratio from EPOS1.99 and QGSJETII-03 at \sqrts~ = 13 and 14 TeV are slightly higher as compared to other models. However, EPOS-LHC, Sibyll2.3d and HIJING model predictions show saturation in the yield ratios at \sqrts~ = 13 and 14 TeV and are in good agreement with the published data at lower energies within statistical fluctuations.

It is important to note that the data from experiments of $K^+/\pi^+$ and $K^-/\pi^-$ ratio as a function of energy confirm the presence of horn which is also observed in HIJING, DPMJETIII and EPOS-LHC models, while rest of the models significantly fails to describe the horn structure. This observed difference may be due to the difference in different models used for current study. In HIJING, Pythia approach to multiple jet processes and nuclear effects for example, jet quenching and parton shadowing is incorporated. The multiple string phenomenological approach exchanges the multiple soft gluons between the quarks or di-quarks present in hadrons which further lead to the longitudinally oriented string-like excitations of those hadrons are also implemented in the HIJING. In order to fix the effect if color flow, valence quarks replaces the flavour of final scattered quark or di-quark. Due to the reason that gluon jets are dominated in $K/\pi$ ratio at intermediate {\ppt} and the ratio is observed to be not sensitive to this~\cite{Sjostrand:1987su, Werner:1988yt, Wang:1991hta}. DPMJETIII, EPOS, QGSJETII-03 and Sibyll2.3d are hadronic interaction models and are based on Gribov Reggeon approach of Pomeron exchange in multiple scatterings. The exchange of individual pomeron occur independently in QGSJETII model which is not true at higher energies where strong overlap of parton cascade exists which further interact with each other. This could be the reason of overprediction of both ratio at higher energies~\cite{Thakuria:2012ie}. Similar to HIJING, Sibyll2.3d also incorporate many concepts of Dual Parton Model. Compared to other models, a pomeron amplitude is not explicitly implemented in Sibyll2.3d. Quantum mechanical model of multiple scattering, EPOS, is based on strings and partons. The production of particles and cross-section observed to be consistent with conservation of energy in EPOS. DPMJETIII, a Hadronic transport model is based on the interactions of strings. The collisions of particles are described through the color exchange and momentum in partons both in target and projectile. This results colorless objects to be joined with these partons which we called ropes, flux tubes or strings.

It is important to mention that the presence of horn in $K^\pm/\pi^\pm$ ratio in experimental measurements at low energy regime is confirmed by various experiments. The hadronic models which confirm the presence of this structure at low energies, two long strings along with valence quark are present at the end and therefore, the contribution of SCET-soft sea quark in resulted particle with low $K^\pm/\pi^\pm$ are already given in Pythia. The fits to cross-section explains the entrance of additional strings at higher energies. The partons in this case are considered like a minijet extension of pQCD events at larger {\ppt} and a continuous transition is expected at cutoff {\ppt} which could results the large {\ppt} and hence larger ratio in case of particles at sea string ends. At reaching energy where new and shorter strings just arrived, there is sizeable fraction of those produced particles which contain string end partons. These strings are then increases and gets longer with the increase in energy and it is expected that the influence of these strings gets weaker which results the decrease in $K^\pm/\pi^\pm$ ratio~\cite{Bhattacharyya:2017rmc}. This case is different in heavy-ion collisions due to the enhancement in the new chains as the results of collision of several nucleons with target and conversely. The newly produced shorter strings results enhancement in $K^\pm/\pi^\pm$ ratio. There is no significant effect from the fusion of strings and rescattering. Due to the presence of uncertainty in parameterization relatd to sea strings, the position of this horn is not a robust prediction.





\begin{table*}[!htb]\label{tab2}
{
\begin{center}
\caption{$K^-/\pi^-$ ratio at different energies, i.e., $\sqrt{s}$ = 6.3 GeV to 14 TeV by using simulations from various MC models along with the comparison of experimental values are displayed here.}
\begin{tabular}{ccccccccccc}\\ \hline\hline

 \small Energy&\small DPMJET-III&\small EPOS1.99 &\small EPOS-LHC  &\small HIJING &\small Sibyll2.3d&\small QGSJET-II &\small Experimental  \\
 $\sqrt{s}$  &      &   &    &   & & \\
\hline
    \small 6.3  & \small 0.031$\pm$0.002 &\small 0.034$\pm$0.006 &\small 0.022$\pm$0.007 &\small 0.044$\pm$0.021 &\small ---- &\small 0.047$\pm$0.009&\small 0.038$\pm$0.01
 \\
    \small 17.3 &\small 0.077$\pm$0.003&\small 0.071$\pm$0.008&\small 0.063$\pm$0.008 &\small 0.090$\pm$0.013 &\small 0.077$\pm$0.015 &\small 0.067$\pm$0.009 &\small 0.092$\pm$0.003
\\
    \small 62.4 &\small 0.097$\pm$0.002&\small 0.095$\pm$0.010 &\small 0.110$\pm$0.009 &\small 0.114$\pm$0.007&\small 0.098$\pm$0.011 &\small 0.087$\pm$0.010 &\small 0.081$\pm$0.005\\
    \small 200 &\small  0.116$\pm$0.005 &\small 0.112$\pm$0.009 &\small 0.108$\pm$0.009&\small 0.113$\pm$0.009 &\small 0.108$\pm$0.009 &\small 0.127$\pm$0.009 &\small 0.102$\pm$0.005
\\
    \small 900 &\small0.113$\pm$0.006&\small 0.127$\pm$0.008 &\small 0.115$\pm$0.008 &\small 0.119$\pm$0.008 &\small 0.117$\pm$0.008 &\small 0.138$\pm$0.007 &\small 0.121$\pm$0.013
\\
    \small 2760 &\small 0.115$\pm$0.006 &\small 0.134$\pm$0.007 &\small 0.118$\pm$0.007 &\small 0.122$\pm$0.006 &\small  0.123$\pm$0.007 &\small 0.142$\pm$0.006 &\small ----\\
    \small 7000 &\small 0.119$\pm$0.007 &\small 0.137$\pm$0.006 &\small 0.121$\pm$0.006 &\small 0.125$\pm$0.008&\small 0.127$\pm$0.006&\small 0.143$\pm$0.006 &\small 0.128$\pm$0.004\\
    \small 13000 &\small ---- &\small 0.139$\pm$0.006 &\small 0.124$\pm$0.006 &\small 0.126$\pm$0.009 &\small 0.130$\pm$0.006 &\small 0.144$\pm$0.005&\small----\\
    \small 14000 &\small---- &\small0.141$\pm$0.006 &\small0.124$\pm$0.006 &\small 0.126$\pm$0.008
 &\small 0.129$\pm$0.006&\small 0.144$\pm$0.005&\small----\\
\hline
\end{tabular}%
\end{center}}
\end{table*}


When comparing Tables 1 and 2, there exists significant difference in both ratios from almost all of the models under study upto \sqrts~= 200 GeV and starts to saturate and hence the no significant difference is observed towards higher energies (\sqrts~= 900 GeV -- 14 TeV). Insignificant difference between the ratios has been observed in case of experimental measurements. This difference may be due to the underlying production mechanisms of $K^+$ and $K^-$. Two possible mechanisms are involved to study the $K^\pm$ production study, pair production and associated production mechanism. Kaon production is heavily influenced by associated production of $s$ and $\bar s$ quark pairs. Since, there is no kaon production through the $\Delta$ channel and hence $K^+$ are produced through $N + N \rightarrow N + X + K^+$ and $\pi + N \rightarrow X + K^+$, where, $X$ is either {\lam} or $\Xi$ hyperon and $N$ is the nucleon. The energy threshold for $N + N \rightarrow N + \Lambda + K$ is significantly lower than thermal production of kaon pairs. The pair production process to produce $K^+$ and $K^-$ is $N + N \rightarrow N + N + K^+ + K^-$. Kaon production in association with a $\Lambda$ is only available to $K^+$ and $K^0$ due to spin degeneracies. The associated production dominates at lower energies and pair production dominates at higher energies in which significantly same number or $K^+$ and $K^-$ are produced. Due to the higher threshold a steeper excitation function of $K^-$ is observed as compared to $K^+$ and therefore, the production cross-section of $K^-$ increases faster at higher energies to that of $K^+$ which results increase in $K^-/\pi^-$ ratio.

\subsection{$K/\pi$ Ratio}

We have also extracted the $K/\pi$ (($K^+ + K^-$)/($\pi^+ + \pi^-$)) yield ratio in $pp$ collisions at various energies using above discussed models. These results are then compared with the $K/\pi$ ratio measured by different experiments~\cite{Pulawski:2015tka, NA49:2009brx, PHENIX:2011rvu, STAR:2008med, ALICE:2011gmo, ALICE:2015ial}. The extracted $K/\pi$ ratio from models as well as experimental values are listed in Table 3.     

\begin{figure}[ht!]
\begin{center}
\includegraphics[width=0.8\textwidth]{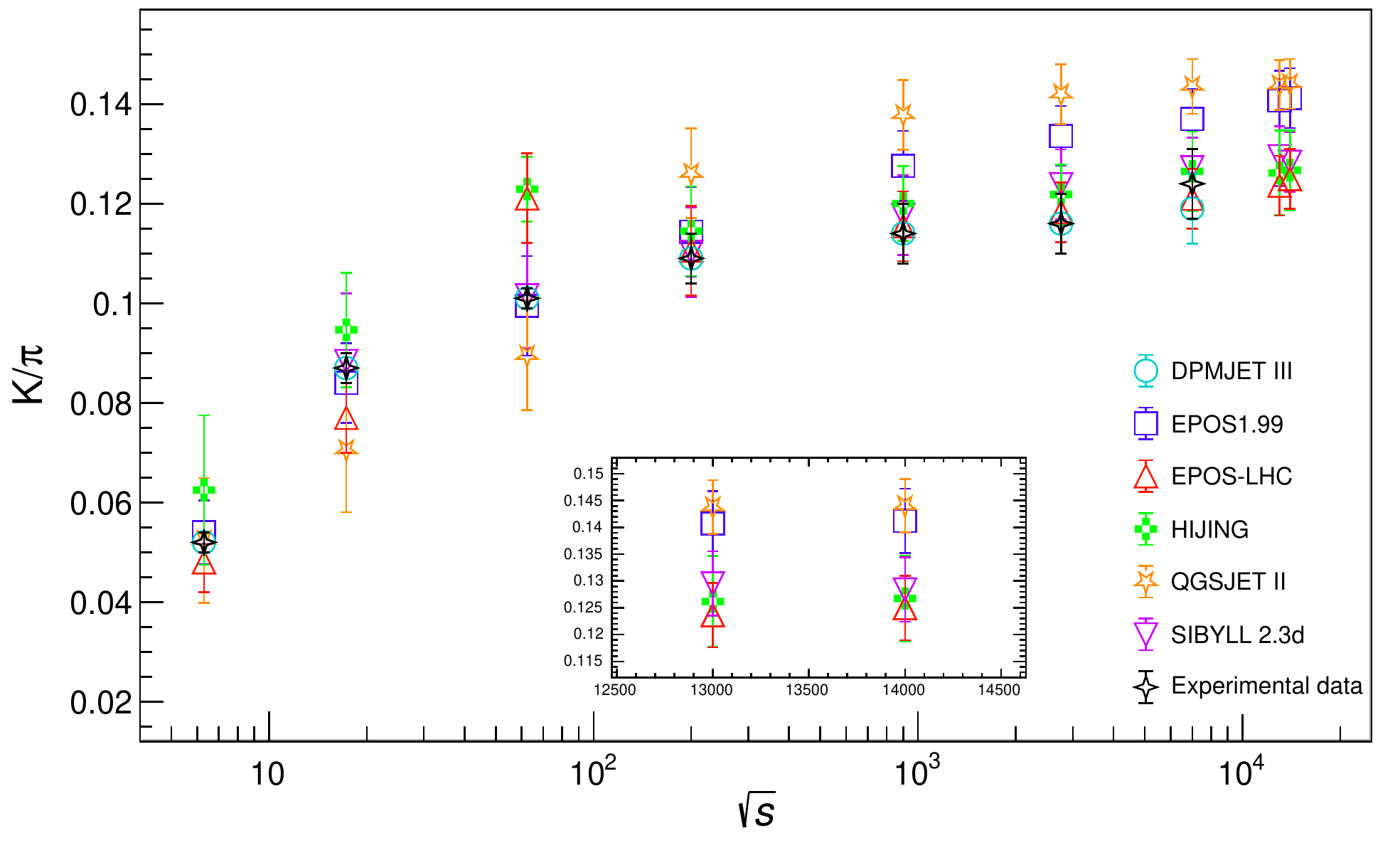}
\caption{ Energy dependence of $K/\pi$ yield ratio at \sqrts~= 6.3 GeV upto \sqrts~= 14 TeV in $pp$ collisions from DPMJETIII, EPOS1.99, EPOS-LHC, HIJING, QGSJETII-03 and Sibyll2.3d compared with the experimental data (where available). The model predictions at \sqrts~ = 13 and 14 TeV are shown in sub-pad for better visualization.}
\label{fig3}
\end{center}
\end{figure}

Figure~\ref{fig3} presents comparison of $K/\pi$ yield ratio from models with experimental measurements. There is smooth increase in the ratio observed in experimental measurements for all energies and almost becomes independent of energy from \sqrts~= 200 GeV. Similar to experimental data, almost all the models starts to saturate from \sqrts~ $>$ 900 GeV. At \sqrts~ $<$ 200 GeV, EPOS1.99 and QGSJETII-03 values are comparable with data while start to over predict at higher energies. Sibyll2.3d, HIJING and EPOS-LHC values of the ratio are in good agreement with experimental data at almost all of energies. While HIJING and EPOS-LHC values of the ratio is almost the same and over predict the data at \sqrts~= 200 GeV. QGSJETII-03 values at \sqrts~ $\le$ 62.4 GeV seems to be lower as compared to experimental data and over estimate the data at higher energies. Overall, the $K/\pi$ from the models are in good agreement with experimental data within statistical fluctuations and saturation is see at higher energies. The predictions of various models at \sqrts~= 13 and 14 TeV suggest that the experimental data should be increasing and lie around the values observed from the model simulations.


\begin{table*}[!htb]\label{tab3}
{
\begin{center}
\caption{$K/\pi$ ratio at different energies, i.e., $\sqrt{s}$ = 6.3 GeV to 14 TeV by using simulations from various MC models along with the comparison of experimental values are displayed here.}
\small

\begin{tabular}{ccccccccccc}\\ \hline\hline
 Energy&DPMJET-III& EPOS1.99 & EPOS-LHC  &HIJING &Sibyll2.3d& QGSJET-II & Experimental  \\
 $\sqrt{s}$  &     & &&&& \\
\hline
    6.3  & 0.052$\pm$0.002 & 0.054$\pm$0.006 & 0.048$\pm$0.006& 0.062$\pm$0.023
 & ---- & 0.045$\pm$0.008&0.052$\pm$0.002\\
    17.3 & 0.087$\pm$0.003 & 0.084$\pm$0.008 & 0.077$\pm$0.007 & 0.095$\pm$0.014& 0.088$\pm$0.013 & 0.070$\pm$0.009 &0.082$\pm$0.003\\
    62.4 & 0.101$\pm$0.002 & 0.099$\pm$0.010 & 0.121$\pm$0.009 & 0.123$\pm$0.009 & 0.101$\pm$0.010 & 0.090$\pm$0.010 &0.094$\pm$0.002\\
    200 & 0.109$\pm$0.005 & 0.114$\pm$0.009 & 0.110$\pm$0.009 & 0.114$\pm$0.009 & 0.110$\pm$0.009 &0.126$\pm$0.009&0.103$\pm$0.008\\
    900 &  0.114$\pm$0.006 & 0.127$\pm$0.007 & 0.115$\pm$0.007 & 0.120$\pm$0.007  &  0.117$\pm$0.007 &0.137$\pm$0.007&0.123$\pm$0.004 \\
    2760 &  0.116$\pm$0.006 & 0.133$\pm$0.006 & 0.118$\pm$0.006 & 0.121$\pm$0.007 & 0.124$\pm$0.007 & 0.142$\pm$0.006&0.124$\pm$0.004\\
    7000 &  0.119$\pm$0.007 & 0.137$\pm$0.006 & 0.121$\pm$0.006 &  0.126$\pm$0.008 & 0.127$\pm$0.006& 0.143$\pm$0.006&0.124$\pm$0.005\\
    13000 & ----&  0.140$\pm$0.006 & 0.123$\pm$0.006 & 0.126$\pm$0.008
 &  0.129$\pm$0.006 & 0.144$\pm$0.005&----\\
    14000 & ---- &0.141$\pm$0.006 & 0.125$\pm$0.006 & 0.126$\pm$0.008 & 0.128$\pm$0.006 & 0.144$\pm$0.005 &----\\
\hline
\end{tabular}%
\end{center}}
\end{table*}


\section{Conclusion}\label{sec4}

A systematic and comprehensive study has been performed to calculate the $K^+/\pi^+$, $K^-/\pi^-$ and $K/\pi$ (($K^+ + K^-$)/($\pi^+ + \pi^-$)) yield ratios as a function of energy in $pp$ collisions at various energies using different model simulations. A good agreement between the model predictions and experimental data has been observed. The difference in both ratios ($K^+/\pi^+$, $K^-/\pi^-$) is observed in experimental measurements and model simulations from \sqrts~= 6.3 GeV to \sqrts~= 200 GeV which becomes insignificant at higher LHC energies. The experimental data suggest the presence of horn-like structure at lower energies. HIJING and EPOS-LHC also suggest similar findings. The production mechanism of kaon plays an important role to study this horn-like structure. There are mainly two mechanisms involved, associated production which is dominated at lower energies and pair production is dominated at higher energies. There may be a possible cross-over between these two mechanisms at energy $\approx$ \sqrts~= 54.4 GeV as reported in Ref.~\cite{Ashraf:2021nkb}. The experimental results of $K/\pi$ (($K^+ + K^-$)/($\pi^+ + \pi^-$)) yield ratio does not show the presence of horn-like structure and smoothly increasing with increasing energies. However, most of the model simulations has similar findings as that of data except HIJING and EPOS-LHC, which slightly over estimate the data at \sqrts~= 62.4 GeV. It has also been observed that the ratio from all model simulations start to saturate at LHC energies. The predictions of these ratios from various model simulations at \sqrts~= 13 and 14 TeV is also studied on the bases of previous available data. These predictions suggest saturation in the yield ratios at higher energies.  

\vskip0.5cm
\textbf{Acknowledgement}\\
The authors would like to show our gratitude to the Ms. Sumaira Ikram from Riphah International University, Mr. Muhammad Salman Ashraf from Institute of Space Technology, Mr. Sudheer Muhammad from Quaid-e-Azam University, Islamabad, Pakistan for sharing their pearls of wisdom with us during the course of this research. 

\vskip0.5cm
\textbf{Availability of Data and Material}\\
The authors declare that all the supported data of this study are available within the article.

\bibliographystyle{unsrt}

\begin{thebibliography}{99}






\bibitem{5}
J.~Rafelski and B.~Muller,
Phys. Rev. Lett. \textbf{48}, 1066 (1982)
[erratum: Phys. Rev. Lett. \textbf{56}, 2334 (1986)]
ssdoi:10.1103/PhysRevLett.48.1066

\bibitem{6}
W.~Florkowski,
``Phenomenology of ultra-relativistic heavy-ion collisions. World Scientific Publishing Company, 2010.''


\bibitem{NA49:2002pzu}
S.~V.~Afanasiev \textit{et al.} [NA49],
Phys. Rev. C \textbf{66} (2002), 054902
doi:10.1103/PhysRevC.66.054902
[arXiv:nucl-ex/0205002 [nucl-ex]].

\bibitem{NA49:2007stj}
C.~Alt \textit{et al.} [NA49],
Phys. Rev. C \textbf{77} (2008), 024903
doi:10.1103/PhysRevC.77.024903
[arXiv:0710.0118 [nucl-ex]].

\bibitem{Pulawski:2015tka}
S.~Pu\l{}awski [NA61],
PoS \textbf{CPOD2014}, 010 (2015)
doi:10.22323/1.217.0010
[arXiv:1502.07916 [nucl-ex]].

\bibitem{Gazdzicki:2010iv}
M.~Gazdzicki, M.~Gorenstein and P.~Seyboth,
Acta Phys. Polon. B \textbf{42} (2011), 307-351
doi:10.5506/APhysPolB.42.307
[arXiv:1006.1765 [hep-ph]].

\bibitem{7}
M.~Gazdzicki and M.~I.~Gorenstein,
Acta Phys. Polon. B \textbf{30} (1999), 2705
[arXiv:hep-ph/9803462 [hep-ph]].




\bibitem{dEnterria:2011twh}
D.~d'Enterria, R.~Engel, T.~Pierog, S.~Ostapchenko and K.~Werner,
Astropart. Phys. \textbf{35}, 98-113 (2011)

\bibitem{Roesler:2000he}
S.~Roesler, R.~Engel and J.~Ranft,
[arXiv:hep-ph/0012252 [hep-ph]].

\bibitem{ATLAS:2020bhl}
G.~Aad \textit{et al.} [ATLAS],
Eur. Phys. J. C \textbf{81}, no.6, 537 (2021)

\bibitem{Bopp:2005cr}
F.~W.~Bopp, J.~Ranft, R.~Engel and S.~Roesler,
Phys. Rev. C \textbf{77}, 014904 (2008)

\bibitem{Pierog:2013ria}
T.~Pierog, I.~Karpenko, J.~M.~Katzy, E.~Yatsenko and K.~Werner,
Phys. Rev. C \textbf{92} (2015) no.3, 034906
doi:10.1103/PhysRevC.92.034906
[arXiv:1306.0121 [hep-ph]].

\bibitem{Sjostrand:2007gs}
T.~Sjostrand, S.~Mrenna and P.~Z.~Skands,
Comput. Phys. Commun. \textbf{178} (2008), 852-867
doi:10.1016/j.cpc.2008.01.036
[arXiv:0710.3820 [hep-ph]].

\bibitem{Bellm:2015jjp}
J.~Bellm, S.~Gieseke, D.~Grellscheid, S.~Pl\"atzer, M.~Rauch, C.~Reuschle, P.~Richardson, P.~Schichtel, M.~H.~Seymour and A.~Si\'odmok, \textit{et al.}
Eur. Phys. J. C \textbf{76} (2016) no.4, 196
doi:10.1140/epjc/s10052-016-4018-8
[arXiv:1512.01178 [hep-ph]].


\bibitem{Pierog:2009zt}
T.~Pierog and K.~Werner,
Nucl. Phys. B Proc. Suppl. \textbf{196}, 102-105 (2009)

\bibitem{Ostapchenko:2007qb}
S.~Ostapchenko,
AIP Conf. Proc. \textbf{928} (2007) no.1, 118-125
doi:10.1063/1.2775904
[arXiv:0706.3784 [hep-ph]].

\bibitem{Engel:2011zzb}
R.~Engel, D.~Heck and T.~Pierog,
Ann. Rev. Nucl. Part. Sci. \textbf{61}, 467-489 (2011)


\bibitem{Kalmykov:1993qe}
N.~N.~Kalmykov and S.~S.~Ostapchenko,
Phys. Atom. Nucl. \textbf{56}, 346-353 (1993)




\bibitem{Ostapchenko:2004ss}
S.~Ostapchenko,
Nucl. Phys. B Proc. Suppl. \textbf{151}, 143-146 (2006)










\bibitem{Wang:1991hta}
X.~N.~Wang and M.~Gyulassy,
Phys. Rev. D \textbf{44}, 3501-3516 (1991)
doi:10.1103/PhysRevD.44.3501


\bibitem{Capella:1979fm}
A.~Capella, U.~Sukhatme and J.~Tran Thanh Van,
Z. Phys. C \textbf{3}, 329-337 (1979)
doi:10.1007/BF01414185

\bibitem{Andersson:2001yu}
B.~Andersson, S.~Mohanty and F.~Soderberg,
Eur. Phys. J. C \textbf{21}, 631-647 (2001)
doi:10.1007/s100520100757
[arXiv:hep-ph/0106185 [hep-ph]].

\bibitem{Riehn:2019jet}
F.~Riehn, R.~Engel, A.~Fedynitch, T.~K.~Gaisser and T.~Stanev,
Phys. Rev. D \textbf{102}, no.6, 063002 (2020)

\bibitem{Riehn:2017mfm}
F.~Riehn, H.~P.~Dembinski, R.~Engel, A.~Fedynitch, T.~K.~Gaisser and T.~Stanev,
PoS \textbf{ICRC2017}, 301 (2018)


\bibitem{CMS:2015zrm}
V.~Khachatryan \textit{et al.} [CMS],
Phys. Lett. B \textbf{751}, 143-163 (2015)







\bibitem{NA49:2009brx}
T.~Anticic \textit{et al.} [NA49],
Eur. Phys. J. C \textbf{65} (2010), 9-63
doi:10.1140/epjc/s10052-009-1172-2
[arXiv:0904.2708 [hep-ex]].



\bibitem{PHENIX:2011rvu}
A.~Adare \textit{et al.} [PHENIX],
Phys. Rev. C \textbf{83} (2011), 064903
doi:10.1103/PhysRevC.83.064903
[arXiv:1102.0753 [nucl-ex]].

\bibitem{STAR:2008med}
B.~I.~Abelev \textit{et al.} [STAR],
Phys. Rev. C \textbf{79} (2009), 034909
doi:10.1103/PhysRevC.79.034909
[arXiv:0808.2041 [nucl-ex]].

\bibitem{ALICE:2011gmo}
K.~Aamodt \textit{et al.} [ALICE],
Eur. Phys. J. C \textbf{71}, 1655 (2011)
doi:10.1140/epjc/s10052-011-1655-9
[arXiv:1101.4110 [hep-ex]].

\bibitem{ALICE:2015ial}
J.~Adam \textit{et al.} [ALICE],
Eur. Phys. J. C \textbf{75} (2015) no.5, 226
doi:10.1140/epjc/s10052-015-3422-9
[arXiv:1504.00024 [nucl-ex]].

\bibitem{Bhattacharyya:2017rmc}
S.~Bhattacharyya, M.~Haiduc, A.~T.~Neagu and E.~Firu,
Adv. High Energy Phys. \textbf{2018}, 6307205 (2018)
doi:10.1155/2018/6307205
[arXiv:1707.07330 [nucl-ex]].

\bibitem{Sjostrand:1987su}
T.~Sjostrand and M.~van Zijl,
Phys. Rev. D \textbf{36}, 2019 (1987)
doi:10.1103/PhysRevD.36.2019

\bibitem{Werner:1988yt}
K.~Werner,
Z. Phys. C \textbf{42}, 85 (1989)
doi:10.1007/BF01565131



\bibitem{Thakuria:2012ie}
C.~C.~Thakuria and K.~Boruah,
[arXiv:1202.3661 [astro-ph.HE]].

\bibitem{Long:2011tk}
H.~Y.~Long, S.~Q.~Feng, D.~M.~Zhou, Y.~L.~Yan, H.~L.~Ma and B.~H.~Sa,
Phys. Rev. C \textbf{84}, 034905 (2011)
doi:10.1103/PhysRevC.84.034905
[arXiv:1103.2618 [hep-ph]].

\bibitem{Ashraf:2021nkb}
M.~U.~Ashraf [STAR],
Nucl. Phys. A \textbf{1005}, 121815 (2021)
doi:10.1016/j.nuclphysa.2020.121815







\end{thebibliography}
%



\end{document}